\documentstyle{aipproc}

\input boxedeps
\SetRokickiEPSFSpecial
\HideDisplacementBoxes

\newcommand{\hepex}[1]{(hep-ex/#1)}
\newcommand{\hepph}[1]{(hep-ph/#1)}
\def\prl#1#2#3{\frenchspacing{\it Phys. Rev. Lett. }{\bf #1}, #2 (19#3)}
\def\pr#1#2#3{\frenchspacing{\it Phys. Rev. D}{\bf #1}, #2 (19#3)}
\def\pl#1#2#3{\frenchspacing{\it Phys. Lett. }{\bf #1}, #2 (19#3)}
\def\np#1#2#3{\frenchspacing{\it Nucl. Phys. }{\bf #1}, #2 (19#3)}

\newcommand{\etal}{{\em et al.}}

\newcommand{\gevcc}{\hbox{ GeV}\!/\!c^2}
\newcommand{\gev}{\hbox{ GeV}}

\newcommand{\mevcc}{\hbox{ MeV}\!/\!c^2}
\newcommand{\tev}{\hbox{ TeV}}
\newcommand{\tevcc}{\hbox{ TeV}\!/\!c^2}

\newcommand{\cm}{\hbox{ cm}}
\newcommand{\pb}{\hbox{ pb}}
\newcommand{\fb}{\hbox{ fb}}

\def\ltap{\mathop{\raisebox{-.4ex}{\rlap{$\sim$}} 
\raisebox{.4ex}{$<$}}}
\def\gtap{\mathop{\raisebox{-.4ex}{\rlap{$\sim$}} 
\raisebox{.4ex}{$>$}}}

\newcommand{\cfrac}[2]{\textstyle \frac{#1}{#2}}

\def\bentarrow{\:\raisebox{1.1ex}{\rlap{$\vert$}}\!\rightarrow}

\def\bothdk#1#2#3#4#5{
	\begin{displaymath}
	\begin{array}{r c l}
	#1 & \rightarrow & #2#3 \\
	 & & \:\raisebox{1.3ex}{\rlap{$\vert$}}\raisebox{-0.5ex}{$\vert$}%
	\phantom{#2}\!\bentarrow #4 \\
	 & & \bentarrow #5
	\end{array}
	\end{displaymath}
		}
\def\twodk#1#2#3#4{
	\begin{displaymath}
	\begin{array}{l}
	#1#2 \\
     \:\raisebox{1.3ex}{\rlap{$\vert$}}\raisebox{-0.5ex}{$\vert$}%
	\phantom{#1}\!\bentarrow #3 \\
	\bentarrow #4
	\end{array}
	\end{displaymath}
		}
\begin{document}
\begin{flushright}
\rule{0pt}{36pt} FERMILAB--CONF--98/059--T 
\vspace{-30pt}
\end{flushright}
\title{The Top Quark and Higgs Boson \\ at Hadron Colliders}

\author{Chris Quigg}
\address{Fermi National Accelerator Laboratory\thanks{Fermilab is 
operated by Universities Research Association Inc.\ under Contract 
No.\ DE-AC02-76CH03000 with the United States Department of Energy.}\\
P.O. Box 500, Batavia, Illinois 60510 USA}

\maketitle

\begin{abstract}
To provide context for discussions of experiments at future muon 
colliders, I survey what is known and what will be known about the 
top quark and the Higgs boson from experiments at hadron colliders.
\end{abstract}

\section*{Introduction}
When we discuss whether there should be muon colliders in our future, 
we must answer a number of important questions.

What machines are possible? When? At what cost?

What are the physics opportunities?

Can we do physics in the environment? 
	(What does it take?)

How will these experiments add to existing knowledge \textit{when they 
	are done?}  The aim of this talk is to provide a survey of what we 
	might expect to know about the top quark and the Higgs boson before 
	a $\mu^{+}\mu^{-}$ collider operates \cite{mahon}.

\section*{The Hadron Colliders}
Let us take a moment to recall the characteristics of the hadron 
colliders that will contribute to the study of the top quark and 
Higgs boson.  The combination of the Fermilab Tevatron and the new 
Main Injector with the CDF and D\O\ detectors will in the future 
bring us $\bar{p}p$ collisions at 2 TeV.  In Fermilab parlance, the 
data now under analysis come from Run I: 100 pb$^{-1}$ at 1.8 TeV,  recorded 
in 1994--1996.  We look forward to the first 2-TeV data.  The approved 
quantum of data is Run II: 2 fb$^{-1}$ in 2000--2002.  Beyond the 
approved running, we are enthusiastic about the physics prospects for 
another high-luminosity run while the Tevatron defines the energy 
frontier.  Although the laboratory hasn't taken a position, we refer 
to this possibility as Run III: 30 fb$^{-1}$ by the year 2006.

On that time scale, the  Large Hadron Collider at CERN will open the 
study of $pp$ collisions at 14 TeV in the ATLAS and CMS detectors.  A 
modest goal for the beginning of the LHC era is to accumulate 
$\int {\mathcal{L}} dt =100\hbox{ fb}^{-1}$ in 2005--2009.

Three elements inform the way we think about experiments in these 
high-energy hadron colliders.  First, they promise high sensitivity from 
high integrated luminosity.  Second, the success of $b$-tagging in the 
hadron-collider environment encourages the hope that heavy-flavor 
tags, and perhaps even triggers, can make future experiments more 
sensitive to the exotic events that may signal new physics.  I have in 
mind here both the CDF Silicon Microvertex Detector (SVX), with resolution $\sim 
	11\mu\hbox{m}$, and the ``soft''-lepton tag used by CDF and D\O\ to 
	identify the transition $b \rightarrow c\ell\nu$.  Third, both the 
	physics and the experimental approach to the new energy regime are 
	colored by the great mass of the top quark.

\section*{The Top Quark}
The top quark has been observed at the Tevatron in the reaction 
\cite{disc}
\bothdk{\bar{p}p}{t}{\;\;\bar{t}+\cdots}{W^{-}\bar{b}}{W^{+}b}
In the Tevatron experiments, the $b$-quarks are identified as 
displaced vertices or through soft-lepton tags.  The channels studied to 
date are dileptons (including $\tau+(e,\mu)$), lepton + 
jets, and all jets.

\subsection*{Top Mass}
The top mass has already been determined to impressive precision.  An 
``unofficial'' average including the latest data from CDF and D\O\ is 
\cite{tmass}
\begin{displaymath}
	m_{t} = 174.3 \pm 5.3 \gevcc \; .
\end{displaymath}

Within the electroweak theory, fermion masses are set by the scale of 
electroweak symmetry breaking $v$ and by apparently arbitrary Yukawa 
couplings, 
\begin{displaymath}
	m_{f} = \frac{\zeta_{f}v}{\sqrt{2}} \approx (176\gevcc)\cdot 
	\zeta_{f}\; .
\end{displaymath}
It is striking that the top quark's Yukawa coupling $\zeta_{t}\approx 1$.
Does this mean that top is special, or might top be the only 
``normal'' fermion, with a mass close to the electroweak scale?

\subsection*{Top Lifetime}
The top-quark lifetime is governed by the semiweak decay $t \rightarrow 
bW^{+}$; the decay width is given by \cite{ttime}
\begin{displaymath}
	\Gamma(t \rightarrow bW^+) = \frac{G_F m_t^3}{8\pi\sqrt{2}}
	|V_{tb}|^{2} \left(1 - \frac{M_{W}^{2}}{m_{t}^{2}}\right)^{\!\!2}
	\left(1 + \frac{2M_{W}^{2}}{m_{t}^{2}}\right) \; .
	\label{tdkapp}
\end{displaymath}
If there are three generations of quarks, so that we can use $3\times 
3$ unitarity to determine $|V_{tb}| = 0.9991 \pm 0.0002 \approx 1$, 
then $\Gamma(t \rightarrow bW^{+}) \approx 1.55\gev$.
This corresponds to a top lifetime, 
\begin{displaymath}
	\tau_{t} \approx 0.4 \times 10^{-24}\hbox{ s} ,
\end{displaymath}
that is very short compared with the time-scale for confinement, 
\begin{displaymath}
	1/\Lambda_{\mathrm{QCD}} \approx \hbox{few}\times 10^{-24}\hbox{ s} .
\end{displaymath}
As a consequence, the top quark decays before it can be hadronized.  
No discrete lines will be observed in the $t\bar{t}$ spectrum, and 
there will be no dressed hadronic states containing top.
This freedom from the confining effects of the strong interaction 
means that the characteristics of top production and the hadronic 
environment near top in phase space should be calculable in 
perturbative QCD.  The fact that top is, in this 
sense, the purest, freest quark we have to study will have important 
consequences for future experiments.

\subsection*{Top Production}
It is useful to summarize some important characteristics of top pair 
production.  At the Tevatron, at 1.8 TeV, the top-pair production 
cross section is $\sigma \approx 6\pb$ \cite{sigcalc}.  Approximately 
90\%  arises 
from the reaction $q\bar{q} \rightarrow t\bar{t}$, and only about 10\% 
from the reaction $gg \rightarrow t\bar{t}$.  Top is a heavy particle 
for the Tevatron, and this is reflected in the dominance of $q\bar{q}$ 
collisions.  The measured cross sections are in reasonable accord with 
this estimate.  CDF measures $7.6^{+1.8}_{-1.5}\pb$ \cite{cdfsig}, while 
D\O\ has determined $5.5\pm 1.8\pb$ \cite{d0sig}.

At the LHC, the pair-production cross section rises to $\sigma \approx 
800\pb$.  The origin of the top events is markedly different.  In 
14-TeV $pp$ collisions, the reaction $q\bar{q} \rightarrow t\bar{t}$ 
accounts for only about 10\% of the rate, whereas $gg \rightarrow 
t\bar{t}$ accounts for 90\%.  At the LHC, top will be a moderately 
light particle.

\subsection*{Future Top Yields}
For Run II, the Tevatron energy will increase to $2\tev$.  
Accordingly, the top-pair production cross section will rise by about 
40\%.  In a run of $30\fb^{-1}$ at 2 TeV, approximately 225K $t\bar{t}$ 
pairs will be produced.  I show in Table \ref{tab:tops} a \textit{Snowmass '96} 
projection of the number of top events available for study in the Tevatron's 
Run II and Run III \cite{frey}.
\begin{table}[h!]
\caption{Anticipated top-quark yields in future Tevatron runs}
\label{tab:tops}
\begin{center}
	\begin{tabular}{lrrc}
		Mode & $2\fb^{-1}$ & $30\fb^{-1}$ & $S/B$  \\
		\hline
		Dilepton & 80 & 1200 & 5:1  \\
		$\ell+ 3\hbox{jets}/1 b$ & 1300 & 20000 & 3:1  \\
		$\ell+ 4\hbox{jets}/2 b$ & 600 & 9000 & 12:1  \\
		Single top (all) & 170 & 2500 & 1:2.2  \\
		Single top ($W^{\star}$) & 20 & 300 & 1:1.3  \\
		\hline
	\end{tabular}
\end{center}
\end{table}
The LHC is a veritable fountain of tops: it will produce 
$8 \times 10^{6}\; t\bar{t}$ pairs in a modest-luminosity exposure of 
only $10\fb^{-1}$.  

It seems reasonable to expect that experiments at the Tevatron and LHC 
will determine the top-quark mass within $\delta m_{t} = (1\hbox{-}2)\gevcc$.

\subsection*{Measuring $|V_{tb}|$}
By studying the number of top events in which they register 0, 1, or 
2 $b$ tags, CDF measures \cite{Vtbmeas} the fraction of top decays 
that lead to $b$ quarks in the final state as
\begin{displaymath}
	B_{b} \equiv \frac{\Gamma(t\rightarrow bW)}{\Gamma(t\rightarrow qW)} =
	\frac{|V_{tb}|^{2}}{|V_{td}|^{2}+|V_{ts}|^{2}+|V_{tb}|^{2}} = 0.99 
	\pm 0.29 \; .
\end{displaymath}
If there are three generations, so that 
$|V_{td}|^{2}+|V_{ts}|^{2}+|V_{tb}|^{2} = 1$, this measurement leads 
to a lower bound on the strength of the $t\bar{b}W$ coupling, 
\begin{displaymath}
	|V_{tb}| > 0.76\;\;(95\%\hbox{ CL}) .
\end{displaymath}
Without the three-generation unitarity constraint, we learn only that
\begin{displaymath}
	|V_{tb}| \gg |V_{td}|, |V_{ts}| .
\end{displaymath}
Increased sensitivity in the forthcoming runs should lead to 
significant improvements in $\delta B_{b}$.  For Run II, we 
anticipate $\pm 10\%$, and for Run III, $\pm$ a few percent.  At the LHC, it 
should be possible to reduce the uncertainty to about $\pm 1\%$.

Direct measurement of the coupling $|V_{tb}|$ will become possible in 
single-top production through the reactions $\bar{q}q \rightarrow 
W^{\star}\rightarrow t\bar{b}$ and $gW \rightarrow t\bar{b}$ \cite{scott}.  The 
cross sections for both reactions are $\propto |V_{tb}|^{2}$.
We can expect to measure the coupling with an uncertainty 
$\delta|V_{tb}| = \pm (10\%, 5\%)$ in Run II and III, using 
both the virtual-$W^{\star}$ channel and $gW$ fusion.  I am not aware 
of any detailed studies for the LHC environment, but the fact that 
the $gW$ fusion cross section is a hundred times larger than at the 
Tevatron means that there will be a very large sample of single-top 
events.
\subsection*{Searches for new physics}
Top decay is an excellent source of 
longitudinally polarized gauge bosons.  In the decay of a massive 
top, $W$-bosons with $|$helicity$|=1$ occur with weight = 1, while 
longitudinally polarized $W$-bosons with helicity = 0 occur with weight 
	$=m_{t}^{2}/M_{W}^{2}$.  If the decays of top proceed by the 
	standard $V\!-\!A$ interaction, we therefore expect that the 
	longitudinal fraction $f_{0} = (m_{t}^{2}/M_{W}^{2})/(1 + 
	m_{t}^{2}/M_{W}^{2}) \approx 70\%$.  The polarization of the 
	$W$-boson is reflected in the decay angular distribution of leptons 
	from its subsequent decay:
\begin{displaymath}
	\frac{d\Gamma(W^+\rightarrow\ell^+\nu_\ell)}{d(\cos\theta)} =
	\cfrac{3}{8}(1-f_0)(1-\cos\theta)^2 + \cfrac{3}{4}f_0 \sin^2\theta 
	\; .
\end{displaymath}
In experiments at the Tevatron, it should be possible to determine the 
longitudinal fraction to $\delta f_{0} = \pm 3\%$ in Run II.  The LHC 
experiments will improve the measurement to $\pm 1 \%$.  Departures 
from the canonical expectation would give a hint of unexpected 
structure at the $t\bar{b}W$ vertex.

The flavor-changing--neutral-current decays
\begin{displaymath}
	 	t \rightarrow 
 	\left(\begin{array}{c}
 		g  \\
 		
 		Z  \\
 		
 		\gamma
 	\end{array}\right) + 
 	\left(\begin{array}{c}
 		c  \\
 		
 		u
 	\end{array}\right)
\end{displaymath}
are unobservably small ($\ll 10^{-10}$) in the standard model 
\cite{fcnc}, but the 
present indirect constraints on the $Zt\bar{c}$ couplings would permit 
branching fractions as large a a few percent \cite{hpz}.  The ultimate 
sensitivity at the Tevatron might reach about 1\% for these decays, 
while the LHC experiments could reach a level of $\sim 10^{-4}$.

It is possible that the rare decay $t \rightarrow bWZ$, with a 
branching fraction $\sim 10^{-6}$ in the standard model, might be 
detectable at the LHC.

Because top is so massive, top decays may surprise by providing a 
conduit to final states that would otherwise be reached with 
difficulty.  One of the favorite targets is the search for a charged 
scalar or pseudoscalar $P^{+}$ in the semiweak decay 
$t \rightarrow bP^{+}$.  Such charged scalars may occur 
in multi-Higgs models, supersymmetry, and technicolor.  Both CDF and 
D\O\ have reported searches \cite{charged}.

\subsection*{Resonances in $t\bar{t}$ Production?}
We have noted that the top quark decays before it can be incorporated 
into a color-singlet hadron.  That fact does not exclude the possibility 
that some new object might include tops among its decay products.  
Because objects associated with the breaking of electroweak symmetry 
tend to couple to fermion mass, the discovery of top opens a new 
window on electroweak symmetry breaking.
Indeed, top-condensate models and multiscale technicolor both imply 
the existence of color-octet resonances with masses of several hundred 
GeV$\!/\!c^{2}$ that decay into $t\bar{t}$.
In technicolor models \cite{eandl}, the prime candidate is a colored pseudoscalar 
produced in the elementary reactions
\begin{displaymath}
	gg \rightarrow \eta_{T} \rightarrow (t\bar{t}, gg) .
\end{displaymath}
Topcolor models \cite{handp} typically include a colored vector state that would 
appear in the reactions
\begin{displaymath}
	q\bar{q} \rightarrow V_{8} \rightarrow (t\bar{t}, b\bar{b}) .
\end{displaymath}
The first hint for such objects would come from the observation of structure 
in the $t\bar{t}$ invariant mass spectrum.  A first look from CDF, based 
on a small sample, resembles the conventional spectrum.

\subsection*{Top-Quark Measurements: Summary}
Until the LHC operates, top-quark measurements will only be possible at 
the Tevatron.  The LHC will, in time, be a prodigious source of tops.  
We expect that the top-quark mass will be determined within $\delta 
m_{t} \approx 1\hbox{-}2\gevcc$ at both the Tevatron and the LHC.  
The production cross section should be measured to $\pm 5\%$ at the Tevatron, 
and to $\pm$ a few \% at the LHC.  The branching fraction
$\delta\Gamma(t\rightarrow bW)/\Gamma(t\rightarrow 
	qW)$ will improve to $\pm 10\%$ in Run II, $\pm$ a few percent in Run 
	III, and $\pm 1\%$ at the LHC.  Studies of single-top production 
	should yield $\delta|V_{tb}| \approx \pm 10\%$ in Run II, and $\pm 5\%$ 
	in Run III at the Tevatron.  In the current Tevatron experiments, 
	searches are under way for $t\bar{t}$ resonances, rare decays, 
	and other signs of new physics.

\section*{The Higgs Boson}
The central challenge in particle physics is to explore the 1-TeV 
scale and elucidate the nature of electroweak symmetry breaking.  A 
key element in this quest is the search for the Higgs boson, the 
agent of electroweak symmetry breaking in the standard electroweak 
theory.  The unique opportunity offered by a muon collider to 
construct a ``Higgs factory'' using the formation reaction
$\mu^{+}\mu^{-} \rightarrow H$
calls attention to a not-too-heavy Higgs boson, as favored in 
supersymmetric models.  In such models, it is plausible that the mass 
of the lightest Higgs boson---which has much in common with the 
standard-model Higgs boson---is no more than $\sim 130\gevcc$.  It is 
important to bear in mind that a heavy Higgs boson remains a logical 
possibility, as we shall see momentarily.  I will abbreviate to the 
search for the standard-model Higgs boson in what follows.
\subsection*{Constraints on the Higgs Mass}
One of the shortcomings of the electroweak theory is that it fails to 
make a prediction for the mass of the Higgs boson.  Perhaps the most 
general statement that can be made is the upper bound derived 
\cite{lqt} from the requirement of perturbative unitarity,
\begin{displaymath}
	M_{H} \ltap \left(\frac{8\pi\sqrt{2}}{3G_{F}}\right)^{\!1/2} \approx 
	1\tevcc\; .
\end{displaymath}
This condition is the most straightforward way to expose the 
importance of the 1-TeV scale.  

We can obtain sharper constraints, in 
the form of upper and lower bounds, at the price of assuming that no 
new physics intervenes up to a cutoff scale $\Lambda$.  The so-called 
``triviality'' bound says that, for a given value of $M_{H}$, the 
electroweak theory makes sense up to a scale \cite{mpp}
\begin{displaymath}
	\Lambda < M_{H}\exp{\left(\frac{4\pi^{2}v^{2}}{3M_{H}^{2}}\right)}\; .
\end{displaymath}
Read the other way, if we regard the electroweak theory as an 
effective theory, apt up to some 
scale $\Lambda$, the triviality bound gives an upper limit on 
$M_{H}$.  If, for example, we demand that the electroweak theory apply 
up to the Planck scale, the Higgs-boson mass must not exceed 
$175\gevcc$.

The requirement that the electroweak vacuum correspond to an absolute 
minimum of the Higgs potential in the face of quantum corrections 
leads to a lower bound, 
\begin{displaymath}
	M_{H}^{2} > \frac{3G_{F}\sqrt{2}}{16\pi^{2}}(2M_{W}^{4}+ M_{Z}^{4}- 
	4m_{t}^{4}) \cdots ,
\end{displaymath}
that also depends on the scale of new physics \cite{linde}.  If we exclude any new 
physics up to the Planck scale, then $M_{H} \gtap 130\gevcc$.

These are informative constraints---given the assumptions that lead to 
them---but they do not really narrow the search.  Crucial guidance 
comes from the direct searches for the standard-model Higgs boson, 
specifically from the study of the reaction $e^{+}e^{-} \rightarrow HZ$ 
at $161+172+183\gev$ in experiments at LEP2.  The four LEP experiments 
examine the $qqbb, \nu\nu qq, \tau\tau qq, \hbox{ and }(ee+\mu\mu)qq$ channels.
Recent running at $\sqrt{s}=183\gev$ is sensitive to Higgs-boson 
masses up to about $82\gevcc$.
Next year's running at $192\gev$ should allow a search up to $M_{H} 
\approx 96\gevcc$.

\subsection*{Clues about $M_{H}$}
Precision electroweak measurements are sensitive to the Higgs-boson 
mass through radiative corrections.  The constraints that arise on 
$M_{H}$ depend on the selection and weighting of the data set and on 
assumptions made about the light-quark contribution to the vacuum polarization 
for $\alpha(M_{Z})$.  I quote three recent analyses by Erler and 
Langacker \cite{erler} to illustrate the range of possibilities.

Including all the precision electroweak data at face value and using a 
selection of measured cross sections for 
$e^{+}e^{-} \rightarrow \hbox{light hadrons}$ to determine 
$\alpha(M_{Z})$, their best fit for the Higgs-boson mass is
$M_{H} = 69^{+85}_{-43}\gevcc$, 
which corresponds to the bounds 
\begin{displaymath}
	M_{H} < \left\{
	\begin{array}{c}
		236  \\
		287  \\
		413
	\end{array}
	  \right\}\gevcc\hbox{ at }\left\{ 
	  \begin{array}{c}
	  	90\%  \\
	  	95\%  \\
	  	99\%
	  \end{array}
	  \right\}\hbox{ CL.}
\end{displaymath}
The central value lies in the range already excluded by direct 
searches for the standard-model Higgs boson.  
Using instead perturbative QCD to compute 
$\delta\alpha_{\mathrm{had}}^{(5)}$, they find a best fit of
$M_{H} = 97^{+79}_{-48}\gevcc$,
which implies the bounds
\begin{displaymath}
	M_{H} < \left\{
	\begin{array}{c}
		229  \\
		273  \\
		377
	\end{array}
	  \right\}\gevcc\hbox{ at }\left\{ 
	  \begin{array}{c}
	  	90\%  \\
	  	95\%  \\
	  	99\%
	  \end{array}
	  \right\}\hbox{ CL.}
\end{displaymath}
In spite of the shift of the central value, the upper bounds are relatively 
stable against the change in $\alpha(M_{Z})$.

However, we may notice that the implications of individual precision 
measurements are not entirely consistent.  For example, SLD's measurement of 
$A_{\mathrm{LR}}$ favors very low---unphysically low---values of $M_{H}$.  
Having no basis to exclude any measurements, one can follow the 
Particle Data Group prescription and rescale the weights of all the 
inconsistent measurements.  Using measured cross sections for 
$e^{+}e^{-} \rightarrow \hbox{light hadrons}$ 
to determine $\alpha(M_{Z})$, Erler and Langacker then find
$M_{H} = 122^{+134}_{-77}\gevcc$,
which leads to the noticeably different bounds
\begin{displaymath}
	M_{H} < \left\{
	\begin{array}{c}
		329  \\
		408  \\
		613
	\end{array}
	  \right\}\gevcc\hbox{ at }\left\{ 
	  \begin{array}{c}
	  	90\%  \\
	  	95\%  \\
	  	99\%
	  \end{array}
	  \right\}\hbox{ CL.}
\end{displaymath}
I have reviewed this work at some length to show the fragility of our 
current estimates of the Higgs-boson mass.  I will nevertheless focus 
on the case of a light Higgs boson, because only a light Higgs boson 
will be accessible at the Tevatron.

\begin{figure}[tb] 
\centerline{\BoxedEPSF{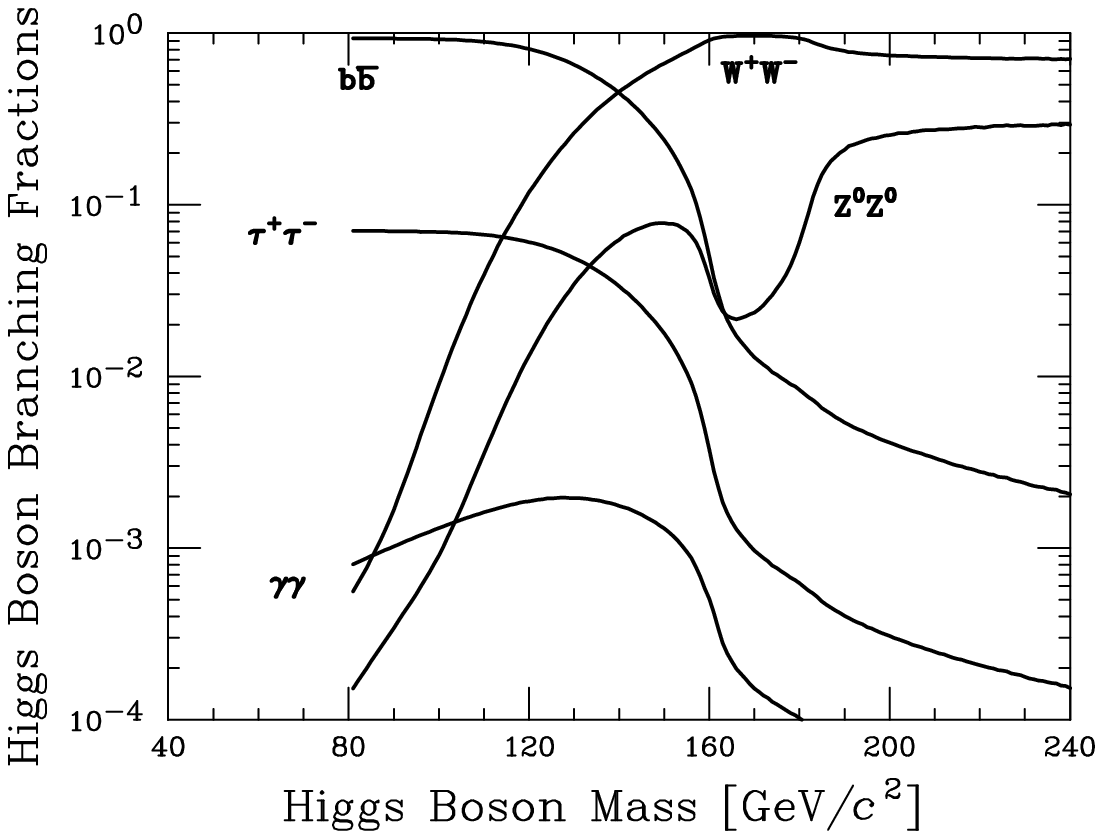  scaled 700}}
\vspace{10pt}
\caption{Branching fractions of a light Higgs boson.}
\label{higgsbr}
\end{figure}

The branching fractions of a light Higgs boson are shown in Figure 
\ref{higgsbr}.  
The most promising channel for searches at the Tevatron will be the 
$b\bar{b}$ mode, for which the branching fraction exceeds about 
50\% throughout the region preferred by supersymmetry and the 
precision electroweak data.

\subsection*{Tevatron Search Strategies}
At the Tevatron, the direct production of a light Higgs boson in 
gluon-gluon fusion $gg \rightarrow H \rightarrow b\bar{b}$ is swamped by 
the ordinary QCD production of $b\bar{b}$ pairs. Even with an 
integrated luminosity of $30\fb^{-1}$, the experiments anticipate 
only $<1\hbox{-}\sigma$ excess, with plausible invariant-mass 
resolution.  It will be possible to calibrate the $b\bar{b}$ mass 
resolution over the region of the Higgs search in Run II: the 
electroweak production of $Z^{0}\rightarrow b\bar{b}$ should stand 
well above background and be observable in Run II.

\begin{figure}[tb] 
\centerline{\BoxedEPSF{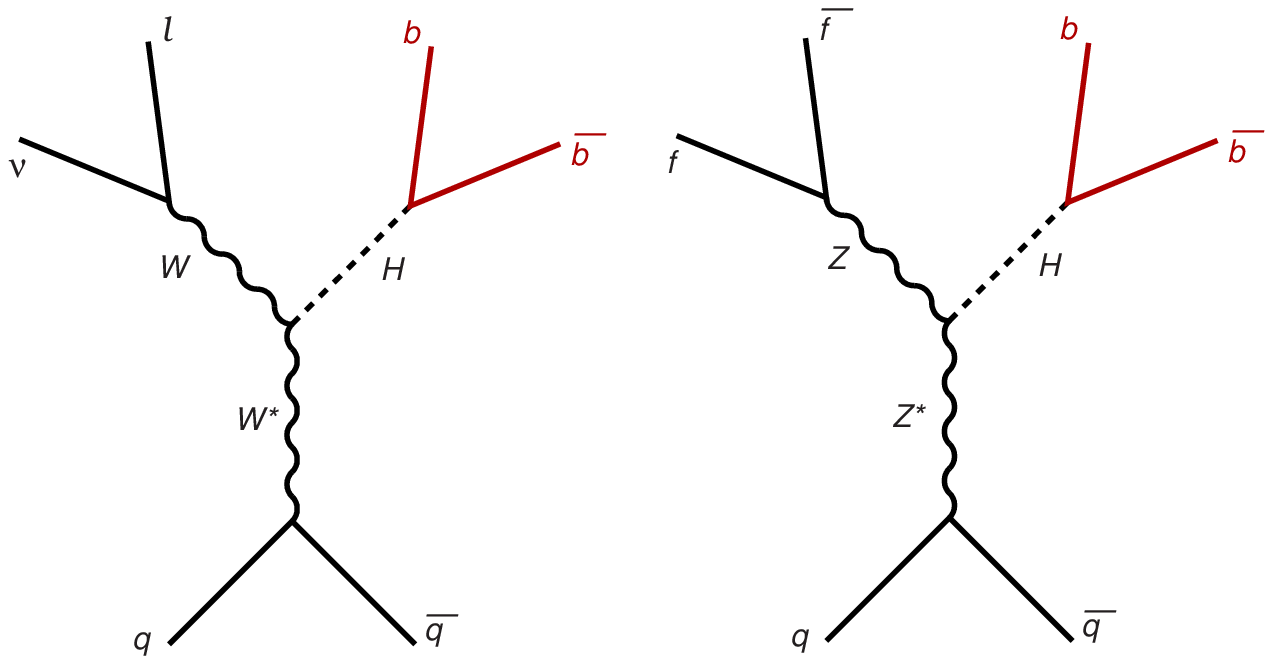  scaled 600}}
\vspace{10pt}
\caption{Feynman diagrams for the associated production of a Higgs boson 
and an electroweak gauge boson.}
\label{assoc}
\end{figure}

The high background in the $b\bar{b}$ channel means that special 
topologies must be employed to improve the ratio of signal to background 
and the significance of an observation.  The high luminosities that can 
be contemplated for a future run argue that the associated-production reactions
    \bothdk{\bar{p}p}{H}{W+\hbox{anything}}{\ell\nu}{b\bar{b}}
and
	\bothdk{\bar{p}p}{H}{Z+\hbox{anything}}{\ell^{+}\ell^{-}+\nu\bar{\nu}}{b\bar{b}}
are plausible candidates for a Higgs discovery at the Tevatron 
\cite{marsw}.  The Feynman diagrams for these processes are shown in 
Figure \ref{assoc}.

The prospects for exploiting these topologies were explored in detail 
in connection with the TeV2000 and TeV33 study groups at Fermilab 
\cite{tevstuds}.  Taking into account what is known, and what might 
conservatively be expected, about sensitivity, mass resolutions, and 
background rejection, these investigations show that it is unlikely 
that a standard-model Higgs boson could be observed in Tevatron Run 
II.  (Note, however, that the ability to use $W\rightarrow q\bar{q}$ 
decays would markedly increase the sensitivity.)  The expected number 
of signal and background events in Run II are collected in Table 
\ref{tab:run2}.
\begin{table}[b!]
\caption{Number of signal and background events in 
Run II ($2\fb^{-1}$) for $WH$ and $ZH$ processes, and 
signal significance \protect\cite{worm}.}
\label{tab:run2}
	\begin{center}
		\begin{tabular}{lcccccccc}
			$M_{H} [\hbox{GeV}\!/\!c^2]$ & 60 & 80 & 90 & 100 & 110 & 120  \\
			\hline
			$WH$ signal $S$ & 45 & 28 &  & 15 &  & 8  \\
			Background $B$ & 139 & 84 &  & 53 &  & 30  \\
			$S/\sqrt{B}$ & 3.8 & 3.1 &  & 2.1 &  & 1.4  \\
			\hline
			$ZH$ signal $S$ &  &  & 7 & 6 & 5 & 3  \\
			Background $B$ &  &  & 36 & 33 & 31 & 25  \\
			$S/\sqrt{B}$ &  &  & 1.2 & 1.1 & 1.0 & 0.7  \\
			\hline
		\end{tabular}
	\end{center}
\end{table}
The prospects are much brighter for Run III.  Indeed, the sensitivity 
to a light Higgs boson is what motivates the integrated luminosity of 
$30\fb^{-1}$ specified for Run III.  The number of events 
projected for Run III, collected in Table \ref{tab:run3}, show that 
a Tevatron experiment could explore the range of Higgs-boson masses up 
to about $125\gevcc$, covering the entire range favored by light-scale 
supersymmetry.
\begin{table}[tb]
\caption{Number of signal and background events in 
Run III ($30\fb^{-1}$) for $WH$ and $ZH$ processes, and 
signal significance \protect\cite{worm}.}
\label{tab:run3}
\begin{center}
	\begin{tabular}{lcccccccc}
		$M_{H} [\hbox{GeV}\!/\!c^2]$ & 60 & 80 & 90 & 100 & 110 & 120  \\
		\hline
		$WH$ signal $S$ & 681 & 420 &  & 228 &  & 117  \\
		Background $B$ & 2085 & 1260 &  & 789 &  & 456  \\
		$S/B$ & 0.33 & 0.33 &  & 0.29 &  & 0.26  \\
		$S/\sqrt{B}$ & 14.9 & 11.8 &  & 8.1 &  & 5.5  \\
		\hline
		$ZH$ signal $S$ &  &  & 108 & 92 & 82 & 51  \\
		Background $B$ &  &  & 533 & 495 & 462 & 378  \\
		$S/B$ &  &  & 0.20 & 0.19 & 0.18 & 0.13  \\
		$S/\sqrt{B}$ &  &  & 4.7 & 4.1 & 3.8 & 2.6  \\
		\hline
	\end{tabular}
\end{center}
\end{table}

\begin{figure}[tb] 
\centerline{\BoxedEPSF{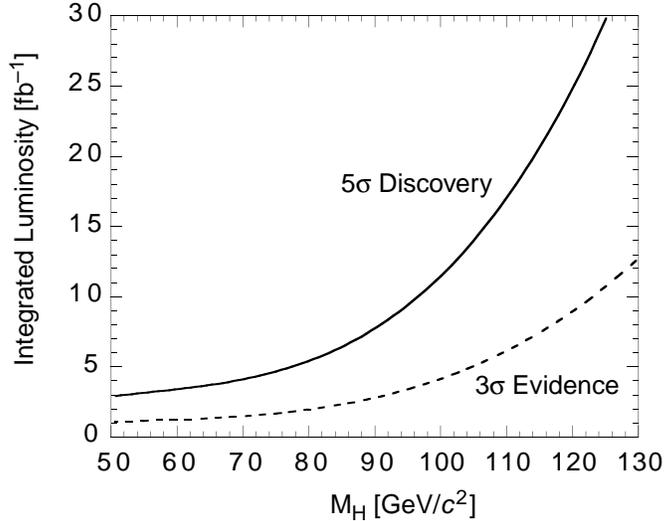  scaled 700}}
\vspace{10pt}
\caption{Luminosity required for the observation of a Higgs boson in 
$WH$ associated production at the Tevatron.}
\label{needed}
\end{figure}

We can make this result a little more transparent by plotting, in 
Figure \ref{needed}, the luminosity needed for a three- or 
five-standard-deviation observation of the Higgs boson at the Tevatron.
We see that, in the $W\!H$ modes discussed, an integrated luminosity of 
$2\fb^{-1}$ is insufficient to detect the standard-model Higgs boson at an 
interesting mass.  About $10\fb^{-1}$ would permit the observation of 
a Higgs boson discovered at LEP2, while $30\fb^{-1}$ would make it 
possible to explore masses up to about $125\gevcc$.  With about 
$10\fb^{-1}$, one could expect a 3-$\sigma$ indication for the Higgs 
boson throughout the low-mass r\'{e}gime.

\begin{figure}[tb] 
\centerline{\BoxedEPSF{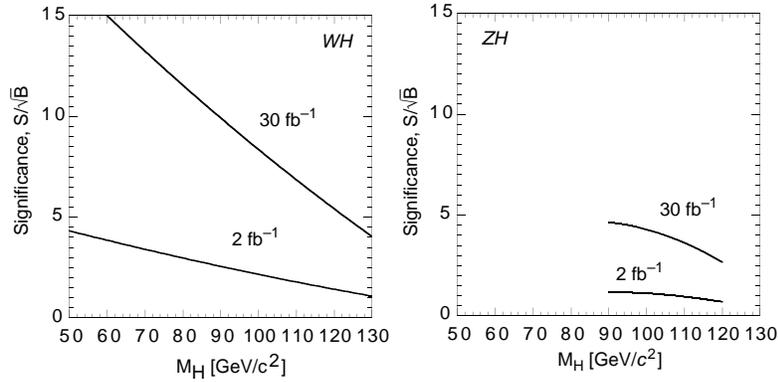  scaled 500}}
\vspace{10pt}
\caption{Significance of Higgs observation in Tevatron Run II \& Run III.}
\label{signif}
\end{figure}

A slightly different cut on the same information is provided in 
Figure \ref{signif}.  There I show the significance of observations in 
the $W\!H$ and $Z\!H$ channels for runs of 2 and  $30\fb^{-1}$.  While 
the $Z\!H$ channel probably would not suffice for an independent 
discovery, it could provide good supporting evidence---and 
complementary measurements---to an observation in $W\!H$.

\subsection*{Higgs at the Tevatron: Summary}
If the Higgs boson is discovered at LEP2, then it should be observed 
at the Tevatron in $W\!H$ with $\int {\mathcal{L}} dt \ltap 
10\fb^{-1}$.  If the Higgs boson lies beyond the reach of LEP2, 
$M_{H} \gtap (95\hbox{-}100)\gevcc$, then a 5-$\sigma$ discovery will 
be possible 
in the $W\!H$ channel in a future Run III of the Tevatron $(30\fb^{-1})$ for 
masses up to $M_{H}\approx 125\gevcc$.  This prospect is the most powerful 
incentive that we have for Run III.  To support this discovery, a
3-$\sigma$ observation will be possible in $Z\!H$ in Run III for masses
up to $M_{H}\approx 110\gevcc$.  In combination, the two observations 
at the Tevatron would imply a $\pm15\%$ measurement of the ratio of 
couplings 	$g^{2}_{W\!W\!H}/g^{2}_{Z\!Z\!H}$.
	If the coupling strength $g_{Z\!Z\!H}$ and the branching fraction
$B(H\rightarrow b\bar{b})$ are known from experiments at LEP2, the 
observations at the Tevatron would make it possible to determine 
$g_{W\!W\!H}$ to $\pm 10\%$.  Over the range of masses accessible at 
the Tevatron, it should be possible to determine the mass of the 
Higgs boson to $\pm (1\hbox{-}3)\gevcc$.

\subsection*{Higgs at the LHC: Summary}
The capabilities of the LHC experiments to search for, and study, the 
Higgs boson are thoroughly documented in the Technical Proposals \cite{tdrs}.  
I will confine myself here to a few summary 
comments.

A 5-$\sigma$ discovery is possible up to $M_{H}\approx 
	800\gevcc$ in a combination of the channels
	\bothdk{H}{Z}{\;Z}{\ell^{+}\ell^{-}}{\ell^{+}\ell^{-},} 
	\twodk{H}{\:W}{\ell\nu}{b\bar{b}}
	and 
\begin{displaymath}
	H \rightarrow \gamma\gamma \hbox{ or perhaps }\tau^{+}\tau^{-}.
\end{displaymath}
The reach of LHC experiments can be extended by making use of the 
channels 
	\bothdk{H}{Z}{\;Z}{\ell^{+}\ell^{-}\hbox{ or }\nu\bar{\nu}}{\hbox{jet 
	jet},}
	and
	\bothdk{H}{W}{\:W}{\ell\nu}{\hbox{jet jet}.}
For Higgs-boson masses below about $300\gevcc$, it should be possible 
to determine the Higgs mass to 100-$300\mevcc$ \cite{gunion}.

\section*{Summary Remarks}
 The Tevatron exists, and will produce important results on the top 
quark and Higgs boson through the next decade.  We can expect 
considerable improvements in the determinations of $m_{t}$ and $M_{W}$, 
as well as increasingly telling searches for nonstandard production and 
decay in Run II ($2\fb^{-1}$).  In the realm of what might be possible 
thereafter, what we have called Run  III ($30\fb^{-1}$) holds great promise 
for refining our knowledge of 	top properties, including the measurement of 
$|V_{tb}|$ in single-top production.  Run III would also extend the search 
for a light Higgs boson throughout the low-mass region favored by 
supersymmetry.  On a related note, if low-scale supersymmetry exists, 
there is every reason to expect that it should be found at the 
Tevatron.

During the week of this workshop, the United States sealed its 
commitment to participate in the construction of the Large Hadron 
Collider at CERN.  The LHC will be a fountain of tops: $\sim 8$ million 
pairs will be produced per year at a luminosity of 
${\mathcal{L}} = 10^{33}\cm^{-2}\hbox{ s}^{-1}$; 
hundreds to thousands of interesting events will be detected each day.  The 
LHC will extend the search for the agent of electroweak symmetry 
	breaking toward $1\tev$.  It will have good sensitivity to the standard-model 
	Higgs boson throughout the interesting range.  The LHC will explore 
	the spectrum of superpartners up to $\sim 1\tevcc$ 
	and make possible detailed measurements of supersymmetric 
	parameters.  Opening a new energy frontier, the LHC will also offer 
	many other possibilities for exploration.
	
\section*{Acknowledgements}
It is a pleasure to thank Dave Cline and his staff for the 
stimulating and pleasant atmosphere of the workshop.  I am grateful 
to Jens Erler, Steve Kuhlmann, Scott Willenbrock, and John Womersley for 
advice and assistance in the preparation of this talk.

\end{document}